\documentclass[a4paper,11pt]{article}

\usepackage{kpfonts}
\usepackage[utf8]{inputenc}
\usepackage[margin=3cm]{geometry}

\usepackage{graphicx}
\usepackage{subcaption}
\usepackage{amsmath, amssymb, amsthm}
\usepackage{mathtools}

\DeclarePairedDelimiter\floor{\lfloor}{\rfloor}
\usepackage{microtype}
\usepackage{booktabs}
\usepackage{enumerate,paralist}
\usepackage{xspace}
\usepackage{comment}
\usepackage{xcolor}
\usepackage[nocompress]{cite}
\usepackage[pdfpagelabels,colorlinks,allcolors=blue]{hyperref}
\usepackage[nameinlink,capitalize,noabbrev]{cleveref}
\hypersetup{
    pdftitle={A Gap in the 42-Queue Layout Algorithm for Planar Graphs},
    pdfauthor={Sergey Pupyrev}
}

\theoremstyle{plain}

\newtheorem{claim}{Claim}
\crefname{claim}{Claim}{Claims}
\Crefname{claim}{Claim}{Claims}

\theoremstyle{definition}  

\graphicspath{{pics/}}

\newcommand{\Qh}{{\ensuremath{\mathcal{Q}}}}

\definecolor{defblue}{rgb}{0.121,0.47,0.705}
\definecolor{defred}{rgb}{0.7,0.37,0.37}
\definecolor{defgreen}{rgb}{0.01,0.65,0.20}

\newcommand{\df}[1]{\textcolor{defblue}{\emph{#1}}}
\newcommand{\alg}[1]{\texttt{#1}\xspace}

\begin{document}

\title{A Gap in the 42-Queue Layout Algorithm for Planar~Graphs}
\author{
    Sergey Pupyrev \\
    spupyrev@gmail.com
}

\date{}
\maketitle

\begin{abstract}
    A \emph{queue layout} of a graph consists of a linear order of the vertices and a partition
    of the edges into queues so that no two edges in a single queue are nested. The minimum number of queues
    needed in a queue layout of a graph is called its \emph{queue number}. The planar product structure theorem states that every planar graph
    is a subgraph of the strong product of a graph of simple treewidth at most $3$, a clique $K_3$, and a path. Such a strong product admits a queue layout with
    $49$ queues (Wood, 2005), which implies
    that the queue number of planar graphs is at most $49$.

    Recently, Bekos, Gronemann, and Raftopoulou (Algorithmica, 2023) investigated how the general approach based on the product structure can be
    optimized for planar graphs. They claim that by appropriately reordering the three vertices in each bag arising from a \emph{tripod},
    it is possible to reduce the queue number of planar graphs to~$42$.
    In this note we highlight a gap in their queue layout algorithm: one of the
    choices required by the algorithm is not guaranteed to exist.  Hence
    the claimed upper bound of $42$ queues is not established by the published proof.
\end{abstract}

\section{Introduction}
Let $G=(V, E)$ be a simple undirected graph with $n = |V|$ and $m = |E|$, and let $\sigma$ be a linear order of the
vertices of $G$. We say that edges $(u, v) \in E$ and $(x, y) \in E$ are \df{nested} with respect to $\sigma$
if $u < x < y < v$ or $x < u < v < y$ in $\sigma$. A set of edges of $G$ forms a \df{queue} with respect to $\sigma$
if no two edges in the set are nested in $\sigma$. Given a linear order $\sigma$ of the vertices of $G$, the edges
$(u_1, v_1), \dots, (u_k, v_k)$ form a \df{rainbow} of size $k$ if $u_1 < \cdots < u_k < v_k < \cdots < v_1$ in $\sigma$.
The edges of $G$ can be partitioned into $k$ queues if and only if there is no rainbow of size $k+1$ in $\sigma$~\cite{HR92}.
A \df{queue layout} of $G$ is a pair $(\sigma,\Qh)$, where $\sigma$ is a linear order of $V$ and $\Qh$ is a partition of $E$
into queues with respect to $\sigma$. The minimum number of queues needed in a queue layout of $G$ is called the
\df{queue number} of $G$.

Queue layouts were introduced by Heath and Rosenberg~\cite{HR92} and have been studied extensively over the
years~\cite{HLR92,BFP13,Pem92,DF15,DW05,DMW05,Wie17,Pup17,ABGKP20,FKMPR23,Katheder0PU25,AlamBGKP22,Pup22}.
Several basic graph classes have bounded queue number.
Every tree has queue number one~\cite{HR92}, outerplanar graphs have queue number at most two~\cite{HLR92}, and series-parallel graphs have queue number at most three~\cite{Wie17}. Alam, Bekos, Gronemann, Kaufmann, and Pupyrev~\cite{ABGKP20} showed that the queue number of
planar $3$-trees is at most five; they also gave a lower bound of four.

The central open problem in the field for many years was whether planar graphs have bounded queue number.
Heath, Leighton, and Rosenberg~\cite{HLR92} conjectured that this is the case.
Di Battista, Frati, and Pach~\cite{BFP13} and Dujmovi{\'c}~\cite{Duj15} obtained logarithmic upper bounds.
The conjecture was settled by Dujmovi{\'c}, Joret, Micek, Morin, Ueckerdt, and Wood~\cite{DJMMUW20}, who proved that every planar graph has queue number at most $49$.
Their proof uses the planar product structure theorem: every planar graph is a subgraph of the strong product of a graph of simple treewidth at most $3$, a clique $K_3$, and a path.
Combining this theorem with Wood's bounds for queue layouts of graph products~\cite{W05} gives the constant $49$.

Bekos, Gronemann, and Raftopoulou~\cite{BGR23} revisited the approach based on the product structure and claimed an improvement from $49$ to $42$.
Their argument uses the freedom to reorder the three vertices associated with each tripod (a bag in the product-structure decomposition).
In this note we point out a gap in their argument: one of the choices required by their ordering rule is not guaranteed to exist.
Consequently, the published proof in~\cite{BGR23} does not establish the claimed upper bound of $42$ queues for planar graphs.

\section{The Product-Structure Approach}
\label{sec:product-structure}

Here we give an overview of the algorithm by Dujmovi{\'c}, Joret, Micek, Morin, Ueckerdt, and Wood~\cite{DJMMUW20},
denoted by \alg{DJMMUW}, and of the refinement by Bekos, Gronemann, and
Raftopoulou~\cite{BGR23}, denoted by \alg{BGR}.

Let $G$ and $H$ be graphs. An \df{$H$-partition} of $G$ is a partition
$\{A_x : x\in V(H)\}$ of $V(G)$ into bags such that, for every edge
$(u,v)\in E(G)$ with $u\in A_x$ and $v\in A_y$, either $x=y$ or
$(x,y)\in E(H)$. In the former case the edge is \df{intra-bag}; in the latter
case it is \df{inter-bag}. A \df{BFS-layering} of $G$ is a partition
$(V_0,V_1,\ldots)$ of $V(G)$ according to the distances from a fixed root.
An edge is \df{intra-layer} if its endpoints lie in the same layer, and
\df{inter-layer} otherwise. The $H$-partition has \df{layered width} $w$
with respect to the BFS-layering if $|A_x\cap V_i|\le w$ for every
$x\in V(H)$ and every layer $V_i$.

\alg{DJMMUW} uses the following general layout lemma. If $H$ has a
$q$-queue layout and $G$ has an $H$-partition of layered width $w$, then
$G$ has a queue layout with $3wq+\floor*{3w/2}$ queues. The vertex order is
built layer by layer. Inside a layer $V_i$, the bags are ordered according to
the vertex order of the $q$-queue layout of $H$: if
$x_1,\ldots,x_h$ is this order, then $V_i$ is ordered as
$A_{x_1}\cap V_i, A_{x_2}\cap V_i,\ldots,A_{x_h}\cap V_i$, with the vertices
inside each set $A_{x_j}\cap V_i$ ordered arbitrarily. The edge assignment
separates five types of edges: intra-layer intra-bag edges, inter-layer
intra-bag edges, intra-layer inter-bag edges, forward inter-layer inter-bag
edges, and backward inter-layer inter-bag edges. Here an inter-layer inter-bag
edge $(u,v)$, with $u\in A_x\cap V_i$ and $v\in A_y\cap V_{i+1}$, is
forward if $x$ precedes $y$ in the vertex order of $H$, and backward
otherwise.

\begin{figure}[!tb]
    \centering
    \includegraphics[width=0.95\textwidth,page=1]{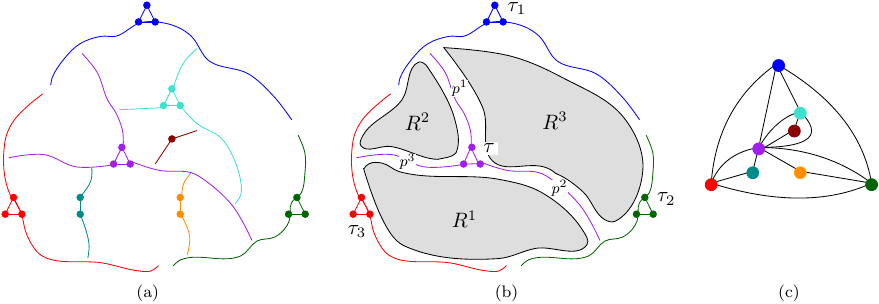}
    \caption{An overview of the tripod decomposition in the planar product structure theorem:
        (a) graph $G$ with tripods;
        (b) a tripod $\tau$ with vertical paths $p^1,p^2,p^3$, and the three recursive regions $R^1,R^2,R^3$ around it;
        (c) the resulting partial planar $3$-tree $H$.}
    \label{fig:overview}
\end{figure}

For planar graphs, the required $H$-partition is obtained from a tripod
decomposition. Assume that $G$ is a plane triangulation, and fix a BFS
tree rooted at a vertex of the outer face. A \df{vertical path} is a path in
this BFS tree. A \df{tripod} is a set of at most three vertical paths whose base
vertices form a triangle of $G$. \alg{DJMMUW} recursively decomposes
regions bounded by three tripods. In a region bounded by
$\tau_1,\tau_2,\tau_3$, Sperner's Lemma gives a new tripod $\tau$, which
splits the region into three smaller regions such that each region is bounded
by at most three tripods out of $\tau,\tau_1,\tau_2,\tau_3$.
The three vertical paths of $\tau$ are denoted by $p^1,p^2,p^3$, and the
three resulting regions by $R^1,R^2,R^3$, where $R^j$ is the region whose
boundary does not contain $p^j$; see \cref{fig:overview}(b).
Graph $H$
has one vertex for each non-empty tripod, and two vertices of $H$ are adjacent
whenever the corresponding tripods are adjacent in $G$. This gives an
$H$-partition of layered width $3$, and $H$ is a partial planar
$3$-tree. Refer to \cref{fig:overview} for an illustration of a tripod decomposition of a plane graph $G$ and the resulting partial planar $3$-tree $H$.

\alg{DJMMUW} then augments $H$ to a maximal planar $3$-tree and applies the
$5$-queue layout of planar $3$-trees by Alam, Bekos, Gronemann, Kaufmann,
and Pupyrev~\cite{ABGKP20}. The planar $3$-tree is decomposed into peeling levels.
Edges whose endpoints lie on the same peeling level are \df{level} edges, and
edges whose endpoints lie on consecutive peeling levels are \df{binding} edges.
The layout uses queues $Q_0,Q_1$ for level edges and queues $Q_2,Q_3,Q_4$
for binding edges. Since the tripod decomposition has layered width $3$ and
$H$ has queue number at most $5$, the general layout lemma
gives $3\cdot 3\cdot 5+\floor*{9/2}=49$ queues.

\alg{BGR} modifies the last step. Its goal is to improve the treatment
of inter-bag edges by observing that every bag has at most three vertices
on each BFS layer, one from each vertical path of the corresponding tripod.
Instead of ordering these three vertices arbitrarily, \alg{BGR} chooses an order of the
three vertical paths of every tripod and uses this order consistently in all BFS
layers. This is the \df{tripod reordering} step, the primary contribution of~\cite{BGR23}.

The crucial observation used in \alg{BGR} is a local separation property for a
new tripod and its parent tripods. Let \(\tau\) be a new tripod with parent
tripods \(\tau_1,\tau_2,\tau_3\); see \cref{fig:overview}(b).
For \(i=1,2,3\), order the three vertical paths of \(\tau_i\) as
\(p_i^1,p_i^2,p_i^3\) so that \(\tau\) lies in the cycle bounded by parts of
\(p_1^1,p_1^2,p_2^1,p_2^2,p_3^1,p_3^2\).
To define the order of tripod vertices, \alg{BGR} needs to carefully augment
the contracted tripod graph $H$ to a maximal planar $3$-tree.
It contracts every tripod to a vertex, keeps the cyclic order of
edges around the contracted vertex, subdivides loops and parallel edges as
needed, and obtains an embedded simple partial $3$-tree $H_0$. It then
completes $H_0$ to a maximal planar $3$-tree $H'$ by a generic
drawing-preserving augmentation~\cite{AP86,KV12}.
Finally, \alg{BGR} applies the $5$-queue layout above to $H'$.
(Here and in the following, we use notation as it is introduced in~\cite{BGR23}.)

It remains to choose the order of the three vertical paths of each tripod
\(\tau\). For the binding queues \(Q_2\) and \(Q_3\) of \(H'\), \alg{BGR}
tries to put first a path missed by a suitable next-level component adjacent
to \(\tau\). Let $v_\tau$ be the vertex of $H'$
corresponding to $\tau$, and suppose that $v_\tau$ lies on peeling level $L_l$.
Among the connected components of the subgraph of $H'$ induced by $L_{l+1}$
that are adjacent to $v_\tau$, consider the components reached by $Q_2$-edges
and by $Q_3$-edges. There are at most two such $Q_2$-components and at most two
such $Q_3$-components. When they exist,
denote them by $c_s^1,c_s^2$ and $c_t^1,c_t^2$, respectively, in the order in
which they appear in the $5$-queue layout. The
\df{tripod-vertices} of such a component $c$, denoted by $T(c)$, are the
vertices of $G$ contained in tripods represented by vertices of $c$; auxiliary
vertices of $H'$ contribute no vertices.
We say that a set of vertices of $G$ \df{touches} a vertical path of $\tau$
if some vertex in the set is adjacent in $G$ to a vertex of the path;
otherwise the set \df{misses} that path.

The tripod reordering starts from the following property of the tripod decomposition.

\begin{claim}[P.6 in~\cite{BGR23}]
    \label{cl:p6}
    There is no edge connecting a vertex of $\tau$ to a vertex of $p_i^3$,
    for $i=1,2,3$.
\end{claim}

Thus, within a single recursive region around a parent tripod, all later
tripods in that region miss the same vertical path of the parent. The proof
in~\cite{BGR23} extends this reasoning to all tripod-vertices in a whole
next-level component and therefore relies on the following property.

\begin{claim}[implicit in Section~4.3 and Lemma~6 of~\cite{BGR23}]
    \label{cl:component-missed}
    For every next-level component \(c\) adjacent to \(v_\tau\), the
    tripod-vertices \(T(c)\) miss a vertical path of \(\tau\).
\end{claim}

The proof of Lemma~6 in~\cite{BGR23} applies \cref{cl:component-missed} to each of
\(c_s^1,c_s^2,c_t^1,c_t^2\). The tripod reordering uses this property to
choose the first vertical path of \(\tau\) so that the relevant second
component misses it. This is recorded in the following property.

\begin{claim}[P.11 and P.12 in~\cite{BGR23}]
    \label{cl:p11-p12}
    If $T(c_s^1)\cup T(c_s^2)$ touches all three vertical paths of $\tau$, then
    $T(c_s^2)$ misses the first vertical path of $\tau$.
    If $T(c_t^1)\cup T(c_t^2)$ touches all three vertical paths of $\tau$, then
    $T(c_t^2)$ misses the first vertical path of $\tau$.
\end{claim}

The tripod reordering in \alg{BGR} uses \cref{cl:p11-p12} to choose the first
vertical path in every tripod; the remaining two vertical paths are ordered arbitrarily.
This determines the vertex order used by \alg{BGR} and leads to the claimed bound of \(42\) queues.
In the next section we show that \cref{cl:component-missed} is false
in general, so no choice of the first vertical path can satisfy
the corresponding statement in \cref{cl:p11-p12}.

\section{A Gap in Tripod Reordering of BGR}

\begin{figure}[!tb]
    \centering
    \captionsetup[subfigure]{justification=centering}
    \begin{subfigure}[b]{.49\linewidth}
        \centering
        \includegraphics[page=1,width=0.99\textwidth]{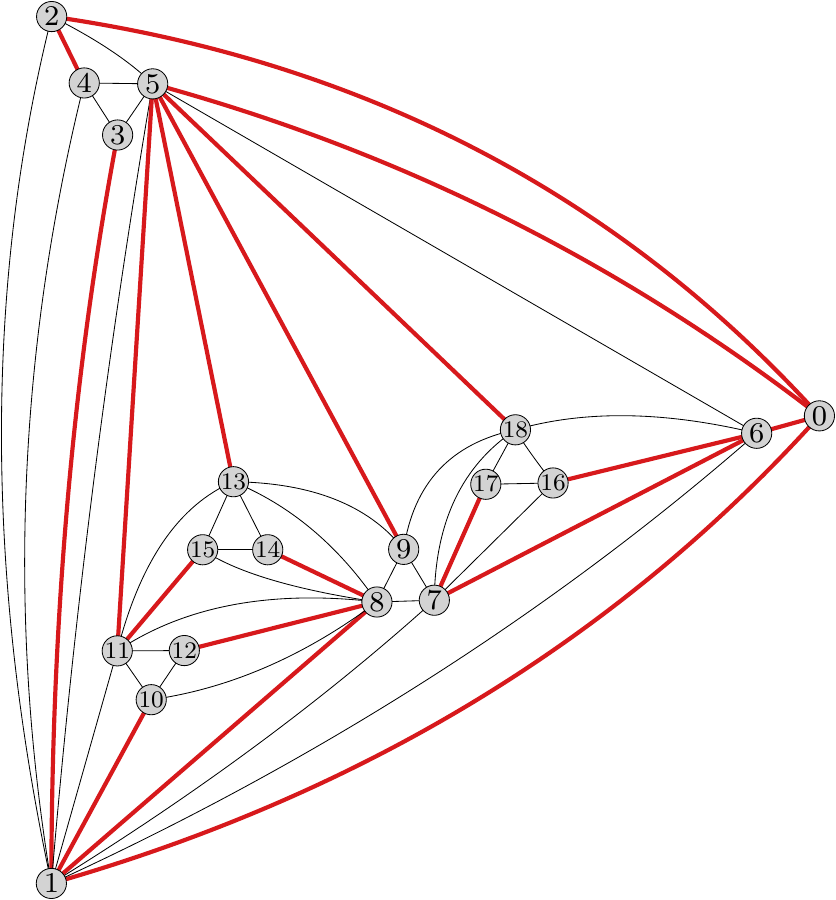}
        \subcaption{Triangulation \(G\)}
        \label{fig:oe1}
    \end{subfigure}
    \begin{subfigure}[b]{.49\linewidth}
        \centering
        \includegraphics[page=3,width=0.99\textwidth]{bgr}
        \subcaption{Tripods in \(G\)}
        \label{fig:oe2}
    \end{subfigure}
    \begin{subfigure}[b]{.49\linewidth}
        \centering
        \includegraphics[page=4,width=0.99\textwidth]{bgr}
        \subcaption{Graph \(H'\)}
        \label{fig:oe3}
    \end{subfigure}
    \begin{subfigure}[b]{.49\linewidth}
        \centering
        \includegraphics[page=5,width=0.99\textwidth]{bgr}
        \subcaption{Component \(c\)}
        \label{fig:oe4}
    \end{subfigure}
    \caption{The counterexample.
        (a) The plane triangulation \(G\), with BFS-tree edges in red.
        (b) The non-empty tripods produced by the recursion.
        (c) The completed graph \(H'\); the vertex \(d\) is auxiliary.
        (d) The next-level component \(c=\{\tau_6,\tau_7,d\}\) adjacent to \(\tau_4\).}
    \label{fig:bgr-cert-g}
\end{figure}

We now give a concrete instance in which \cref{cl:component-missed} is false.
Consequently, the tripod reordering using \cref{cl:p11-p12} cannot be applied.
The graph, its tripod decomposition, and
the relevant component of \(H'\) are shown in \cref{fig:bgr-cert-g}.

\subsection{The Instance}

Let \(G\) be the plane triangulation with $n=19$ vertices shown in \cref{fig:oe1}. The outer face is \((0,1,2)\), and the red edges in
the figure form the BFS tree rooted at \(0\).

The \alg{BGR} recursion starts with the boundary tripods
\(\tau_0=(0)\), \(\tau_1=(1)\), and \(\tau_2=(2)\).  It then finds the
Sperner face \((3,4,5)\), which creates the tripod \(\tau_3\).
In the region with boundary
\([5,0,1,3]\), it finds the Sperner face \((7,8,9)\), creating
\(\tau_4\) with vertical paths \(p_4^1=(6,7)\), \(p_4^2=(8)\), and \(p_4^3=(9)\).  The recursion
then creates \(\tau_5\) from the face \((10,11,12)\), \(\tau_6\) from the face
\((13,14,15)\), and \(\tau_7\) from the face \((16,17,18)\), as shown in
\cref{fig:oe2}. Observe that \cref{cl:p6} holds for each constructed
tripod; for example,
\(V(\tau_6)\) touches \(p_4^2\) and \(p_4^3\), and misses \(p_4^1\), while
\(V(\tau_7)\) touches \(p_4^1\) and \(p_4^3\), and misses \(p_4^2\).

\alg{BGR} continues by building $H_0$ and augmenting it to a maximal planar
3-tree, $H'$. For this instance, $H_0$ has parallel edges between the vertices
corresponding to $\tau_3$ and $\tau_4$. One of the two edges is subdivided by a vertex $d$, and
the graph is augmented to $H'$ by adding edges $(\tau_6, d)$ and $(\tau_7, d)$;
see \cref{fig:oe3}. We stress that \alg{BGR} does not prescribe
which of the two parallel edges is subdivided, leaving the choice to an implementation.
The following elimination order certifies that \(H'\) is a maximal planar
3-tree with the embedding shown in \cref{fig:oe3}:
\(
\tau_7,\ d,\ \tau_6,\ \tau_5,\ \tau_4,\ \tau_3,
\)
leaving the outer triangle \((\tau_0,\tau_1,\tau_2)\).

The next step of \alg{BGR} is to construct a $5$-queue layout of $H'$ following
an approach of~\cite{ABGKP20}.
The peeling levels of \(H'\), rooted at the outer triangle
\(\{\tau_0,\tau_1,\tau_2\}\), are
\(L_0=\{\tau_0,\tau_1,\tau_2\}\),
\(L_1=\{\tau_3,\tau_4,\tau_5\}\), and
\(L_2=\{\tau_6,\tau_7,d\}\).
The \(L_2\)-neighbors of \(\tau_4\) relevant to the tripod reordering step all
belong to the same connected component \(c=\{\tau_6,\tau_7,d\}\).
The queue assigned to the edges between \(\tau_4\) and \(c\) depends on the
choice of the base edge in the 5-queue layout of \(H'\). Since \alg{BGR} does
not prescribe this choice, there are valid executions in which \(c\) is one of
the \(Q_2\)-components or one of the \(Q_3\)-components. We consider such an
execution, so \cref{cl:component-missed} must apply to \(c\).

\subsection{The Failure}

Since the auxiliary vertex \(d\) contributes no vertices of \(G\), we have
\(T(c)=V(\tau_6)\cup V(\tau_7)\).
As observed above, \(V(\tau_6)\) misses \(p_4^1\), while \(V(\tau_7)\) misses
\(p_4^2\). These are different paths of \(\tau_4\), and their union touches all
three vertical paths \(p_4^1,p_4^2,p_4^3\).
Hence \(T(c)\) misses no vertical path of \(\tau_4\), contradicting
\cref{cl:component-missed}. The proof of Lemma~6 in~\cite{BGR23} applies
\cref{cl:component-missed} to every \(Q_2\)- and \(Q_3\)-component, including
\(c\), and therefore does not go through. Furthermore, if \(c\) is the
component used by the tripod reordering to choose the first path, then the
corresponding statement in \cref{cl:p11-p12} is also false.

The core problem in the correctness proof of \alg{BGR}~\cite{BGR23} is the step that applies \cref{cl:p6} to a whole next-level component of \(H'\).
\Cref{cl:p6} is valid for tripods created in a
single recursive region,
but the augmentation of \(H_0\) to \(H'\) may create a next-level component
containing tripods from different recursive regions around the same tripod.
Each region supplies a missed path, yet
the missed paths need not coincide. Consequently, after taking the union of
tripod-vertices over the component, no missed path need remain.

This does not imply that \(G\) has queue number greater than \(42\); it shows
that the published construction does not produce the asserted layout from the
stated argument.

\section{Discussion}

Here we discuss two possible directions for correcting the \alg{BGR} algorithm
and highlight associated challenges.

The first approach is to strengthen the augmentation step used by \alg{BGR}.
Instead of completing \(H_0\) to a maximal planar 3-tree \(H'\) arbitrarily,
one could try to ensure that every component of the next peeling level adjacent to a tripod remains contained in a single recursive region around that tripod.
Then the local separation property (\cref{cl:p6}) would imply the global missed-path property (\cref{cl:component-missed}) required for the tripod
reordering step.
This approach amounts to a constrained, region-preserving augmentation of a partial planar 3-tree, where a naive generic construction~\cite{AP86,KV12} does not suffice.

The second approach is to avoid the augmentation step altogether.
Contracting tripods naturally gives an embedded non-simple graph, with loops and parallel edges encoding different ports or sectors around the same tripod. One could try to extend the 5-queue algorithm for planar 3-trees
of Alam, Bekos, Gronemann, Kaufmann, and Pupyrev~\cite{ABGKP20} to multigraphs.
The main obstacle is that the 5-queue algorithm relies on maximal plane 3-trees: each next-level component lies in a triangular face of the previous level and has three distinct previous-level neighbors that determine the binding queues.
For an embedded multigraph, faces may be triangular walks with repeated vertices, loops, or parallel edges, so the roles used to assign the binding queues are no longer well defined.

Finally, we emphasize that the upper bound of $42$ in~\cite{BGR23} combines two
separate ideas for improving the original $49$ bound of \alg{DJMMUW}~\cite{DJMMUW20}.
The $6$-queue saving for inter-bag edges proved in Lemma~6 is not established by the published proof, as shown above.
A separate observation, also mentioned originally in \cite{DJMMUW20}, saves $1$ queue for intra-bag inter-layer edges.
Therefore, the published proof establishes only the upper bound of $48$ queues
for planar graphs.

\bibliographystyle{abbrv}
\bibliography{refs}

@article{DF15,
	author    = {Vida Dujmovi{\'c} and
	Fabrizio Frati},
	title     = {Stack and Queue Layouts via Layered Separators},
	journal   = {J. Graph Algorithms Appl.},
	volume    = {22},
	number    = {1},
	pages     = {89--99},
	year      = {2018},
}

@article{Duj15,
    title={Graph layouts via layered separators},
    author={Dujmovi{\'c}, Vida},
    journal={Journal of Combinatorial Theory, Series B},
    volume={110},
    pages={79--89},
    year={2015},
    publisher={Elsevier}
}

@article{HR92,
    title={Laying out graphs using queues},
    author={Heath, Lenwood S and Rosenberg, Arnold L},
    journal={SIAM Journal on Computing},
    volume={21},
    number={5},
    pages={927--958},
    year={1992},
    publisher={SIAM}
}

@article{HLR92,
    title={Comparing queues and stacks as machines for laying out graphs},
    author={Heath, Lenwood S and Leighton, Frank Thomson and Rosenberg, Arnold L},
    journal={SIAM Journal on Discrete Mathematics},
    volume={5},
    number={3},
    pages={398--412},
    year={1992},
    publisher={SIAM}
}

@article{Wie17,
    author    = {Veit Wiechert},
    title     = {On the Queue-Number of Graphs with Bounded Tree-Width},
    journal   = {Electr. J. Comb.},
    volume    = {24},
    number    = {1},
    pages     = {P1.65},
    year      = {2017},
}

@article{DMW05,
    title={Layout of graphs with bounded tree-width},
    author={Dujmovi{\'c}, Vida and Morin, Pat and Wood, David R},
    journal={SIAM Journal on Computing},
    volume={34},
    number={3},
    pages={553--579},
    year={2005},
    publisher={SIAM}
}

@phdthesis{Pem92,
    title={Exploring the powers of stacks and queues via graph layouts},
    author={Pemmaraju, Sriram V},
    year={1992},
    school={Virginia Tech}
}

@article{DW05,
    title={Stacks, Queues and Tracks: Layouts of Graph Subdivisions},
    author={Dujmovi{\'c}, Vida and Wood, David R},
    journal={Discrete Mathematics and Theoretical Computer Science},
    volume={7},
    pages={155--202},
    year={2005}
}

@article{BFP13,
    title={On the queue number of planar graphs},
    author={Di Battista, Giuseppe and Frati, Fabrizio and Pach, Janos},
    journal={SIAM Journal on Computing},
    volume={42},
    number={6},
    pages={2243--2285},
    year={2013},
    publisher={SIAM}
}

@article{ABGKP20,
    author       = {Jawaherul Md. Alam and
    Michael A. Bekos and
    Martin Gronemann and
    Michael Kaufmann and
    Sergey Pupyrev},
    title        = {Queue Layouts of Planar 3-Trees},
    journal      = {Algorithmica},
    volume       = {82},
    number       = {9},
    pages        = {2564--2585},
    year         = {2020},
    doi          = {10.1007/S00453-020-00697-4},
}

@inproceedings{Pup17,
	author    = {Sergey Pupyrev},
	title     = {Mixed Linear Layouts of Planar Graphs},
	booktitle = {International Symposium on Graph Drawing and Network Visualization},
	series    = {LNCS},
	volume    = {10692},
	pages     = {197--209},
	publisher = {Springer},
	year      = {2017}
}

@article{DJMMUW20,
    title={Planar graphs have bounded queue-number},
    author={Dujmovi{\'c}, Vida and Joret, Gwena{\"e}l and Micek, Piotr and Morin, Pat and Ueckerdt, Torsten and Wood, David R},
    journal={Journal of the ACM},
    volume={67},
    number={4},
    pages={1--38},
    year={2020},
    publisher={ACM New York, NY, USA},
    doi = {10.1145/3385731}
}

@article{BGR23,
    title={An improved upper bound on the queue number of planar graphs},
    author={Bekos, Michael A and Gronemann, Martin and Raftopoulou, Chrysanthi},
    journal={Algorithmica},
    volume={85},
    number={2},
    pages={544--562},
    year={2023},
    publisher={Springer}
}

@article{W05,
    title={Queue layouts of graph products and powers},
    author={Wood, David R},
    journal={Discrete Mathematics \& Theoretical Computer Science},
    volume={7},
    year={2005},
    publisher={Episciences.org}
}

@InProceedings{FKMPR23,
    author="F{\"o}rster, Henry
    and Kaufmann, Michael
    and Merker, Laura
    and Pupyrev, Sergey
    and Raftopoulou, Chrysanthi",
    editor="Morin, Pat and Suri, Subhash",
    title="Linear Layouts of Bipartite Planar Graphs",
    booktitle="Algorithms and Data Structures",
    year="2023",
    publisher="Springer Nature Switzerland",
    address="Cham",
    pages="444--459",
}

@inproceedings{Katheder0PU25,
    author       = {Julia Katheder and
    Michael Kaufmann and
    Sergey Pupyrev and
    Torsten Ueckerdt},
    title        = {Transforming Stacks into Queues: Mixed and Separated Layouts of Graphs},
    booktitle    = {STACS},
    series       = {LIPIcs},
    volume       = {327},
    pages        = {56:1--56:18},
    publisher    = {Schloss Dagstuhl - Leibniz-Zentrum f{\"{u}}r Informatik},
    year         = {2025},
    doi          = {10.4230/LIPICS.STACS.2025.56},
}

@article{AlamBGKP22,
    author       = {Jawaherul Md. Alam and
    Michael A. Bekos and
    Martin Gronemann and
    Michael Kaufmann and
    Sergey Pupyrev},
    title        = {The mixed page number of graphs},
    journal      = {Theor. Comput. Sci.},
    volume       = {931},
    pages        = {131--141},
    year         = {2022},
    doi          = {10.1016/J.TCS.2022.07.036},
}

@article{KV12,
    author       = {Jan Kratochv{\'{\i}}l and Michal Vaner},
    title        = {A note on planar partial 3-trees},
    journal      = {CoRR},
    volume       = {abs/1210.8113},
    year         = {2012},
    url          = {http://arxiv.org/abs/1210.8113},
    eprinttype   = {arXiv},
    eprint       = {1210.8113},
}

@article{AP86,
    title={Characterization and recognition of partial 3-trees},
    author={Arnborg, Stefan and Proskurowski, Andrzej},
    journal={SIAM Journal on Algebraic Discrete Methods},
    volume={7},
    number={2},
    pages={305--314},
    year={1986},
    publisher={SIAM}
}

@inproceedings{Pup22,
    author       = {Sergey Pupyrev},
    editor       = {Patrizio Angelini and Reinhard von Hanxleden},
    title        = {Queue Layouts of Two-Dimensional Posets},
    booktitle    = {International Symposium on Graph Drawing and Network Visualization},
    series       = {LNCS},
    volume       = {13764},
    pages        = {353--360},
    publisher    = {Springer},
    year         = {2022},
    doi          = {10.1007/978-3-031-22203-0\_25},
}

\end{document}